\documentclass[prl,aps,twocolumn,showpacs,floatfix,10pt]{revtex4-1}
\usepackage{appendix}
\usepackage{epsfig}
\usepackage{amssymb}
\usepackage{graphicx,color}

\newcommand {\be} {\begin{equation}}
\newcommand {\ee} {\end{equation}}
\newcommand {\dm} {$\partial M/\partial n$}
\newcommand {\dchi} {$\partial \chi/\partial n$}

\begin{document}
\preprint{APS/123-QED}
\title{Spin-Droplet State of an Interacting 2D Electron System}

\author{N.~Teneh$^{1}$, A.~Yu.~Kuntsevich$^{2}$, V.~M.~Pudalov$^{2,3}$ and M.~Reznikov$^{1}$}
\affiliation{
$^{1}$Solid State Institute, Technion, Haifa 32000, Israel\\
$^{2}$ P.N. Lebedev Physical Institute, Moscow, 119991 Russia \\
$^3$  Moscow Institute of Physics and Technology, Moscow, 141700, Russia }

\begin{abstract}
We report thermodynamic magnetization measurements of two-dimensional electrons in several high mobility Si metal-oxide-semiconductor field-effect transistors. We provide evidence for an easily polarizable electron state in a wide density range from insulating to deep into the metallic phase. The temperature and magnetic field dependence of the magnetization is consistent with the formation of large-spin droplets in the insulating phase. These droplets melt in the metallic phase with increasing density and temperature, although they survive up to large densities.
\end{abstract}

\date{\today}
\maketitle

Magnetic ordering of a low-density electron system is determined by the interplay between the electronic Coulomb interaction and Pauli principle. As the density decreases, the ratio between the interaction and Fermi energies increases, pushing the system towards a ferromagnetic instability.

In the Hartree-Fock approximation, the Bloch instability, a first-order transition from unpolarized to fully polarized state, happens at an unrealistically small $r_s\approx 2$.  In the opposite limit of short-range interaction the Stoner instability, a second-order phase transition characterized by divergent spin susceptibility, is expected. The hierarchy of these transitions is discussed in Ref.\,\cite{Zhang2005} within the RPA approximation. Numerical simulations for a clean single-valley two-dimensional electron system (2DES)\,\cite{Attaccalite02} predict a Bloch instability at $r_s\approx 25$ followed by Wigner crystallization \cite{Tanatar_1989} at $r_s \approx 37$.

However, at very low densities a realistic system cannot be treated as a clean one: even small potential fluctuations due to inevitably present disorder become dominant and lead to Anderson localization. At higher densities intricate interplay between disorder and interactions manifests itself as a metal-insulator transition (MIT) at some density $n_c$\,\footnote{$n_c$ is the density below which the metallic temperature dependence of resistivity switches to the insulating one. We do not discuss here whether a true conductive state exists at zero temperature}. Experimental observations\,\cite{Ilani01} and theoretical arguments\,\cite{Tripathi2006, Shi2002} suggest that a 2DES becomes strongly nonuniform at densities lower than $n_c$; for $n < n_c$ a 2DES can be considered as consisting of weakly coupled disordered quantum dots. Disorder is also expected to drive a 2D system further towards ferromagnetic instability\,\cite{Fleury2010,DePalo2009}. In particular, a disordered quantum dot is predicted to have a finite spin in the ground state, a phenomenon analogous to the Stoner instability\,\cite{AndreevKamenev,Shep2001}. Experiments on quantum dots\,\cite{Ghosh2004} in GaAs indeed found spontaneous spin polarization at $r_s\sim 7.6$, much smaller than the expected value for a clean system\,\cite{Attaccalite02}.

Coulomb interactions lead to renormalization of the Fermi-liquid constants, notably the density of states and the effective $g$-factor, $g=g_0/(1+F_0^\sigma)$ with Stoner instability expected at $F_0^\sigma=-1$. Negative $F_0^\sigma$ can indeed be deduced for different 2DESs from measurements of Shubnikov-de Haas (ShdH) oscillations \,\cite{Pudalovsdh02,Tutuc2002} and the temperature-dependent resistivity\,\cite{Kvon02,Proskuryakov02,Klimov08}. Scaling analysis of the magnetoresistance led the authors of Refs.\,\cite{SKDK,VZ2} to suggest a quantum phase transition into a ferromagnetic state at $n_c$. This conjecture was contested in Refs.\,\cite{Prus2003} and\,\cite{Pudalovsdh02,*PGK2001,Tutuc2002,Zhu2003} on the basis of thermodynamic and ShdH measurements, respectively. It should be emphasized that all measurements have been done at relatively high magnetic fields, at which Zeeman splitting exceeds the temperature ($g^*\mu_B B >k_BT$); apparently for this reason no anticipated\,\cite{Finkelstein84,Castellani84,Punnoose2005} divergency  of the 2D spin susceptibility with decreasing temperature has been revealed experimentally until now.

We report an observation of a spin-droplet state in low disorder Si metal-oxide-semiconductor field-effect transistors (MOSFETs) on the basis of thermodynamic magnetization measurements. We used a recharging technique suggested in Ref.\,\cite{Pudalov_86,*Pudalov_85} for energy spectrum reconstruction and developed for magnetization measurements in Ref.\,\cite{Prus2003}. In this technique, the recharging current between the gate and the 2DES flows in response to a change of the 2DES chemical potential $\mu$ caused by modulation of the in-plane magnetic field at a constant gate voltage\,\cite{Reznikov_2010}
\begin{equation}\label{dmudb}
\frac{e^2}{\tilde c}\frac{d n}{d B}= -\frac{\partial \mu}{\partial B} + \frac{e^2n}{c_0^2}\frac{\partial c_0}{\partial
B}
\end{equation}
where $n$ is the electron density, $c_0=\epsilon/4\pi d$ is the geometrical capacitance  per unit area, and the capacitance $\tilde c=(c_0^{-1}+e^2\partial n/\partial\mu)^{-1}$ includes compressibility contribution. Note that, as long as the {\it geometrical} capacitance is magnetic field independent, the recharging current is proportional to ${\partial\mu}/{\partial B}$, and therefore the recharging method is distinct from the magneto-capacitance one. We used Si MOSFETs with $\approx 200 {\rm nm}$ gate oxide, much thicker than the 2DES $\approx 4 {\rm nm}$ and thus $c_0$ is set by the gate oxide and is almost magnetic field independent\,\cite{Reznikov_2010}.

Importantly, Eq.\,(\ref{dmudb}) holds even when the capacitance $\tilde c$ acquires an imaginary part due to contact and channel resistances~\cite{Reznikov_2010}. This allowed us to extend the measurements deep into the insulating phase, with the only constraint being an ability to accurately measure recharging current $I$ and capacitance $\tilde c$ at the frequency $\omega$ of magnetic field modulation. In practice, the sample resistance was required to be below $\approx 100\,{\rm M\Omega}$, which happened, e.g., at less than half of the critical density $n_c$ at 4 K. The recharging current per unit area is then given by $I=(-i\omega \tilde c/e)({\partial \mu}/{\partial B})\delta B$, where $\delta B$ is the modulation depth. By virtue of the Maxwell relation $\partial\mu/\partial B=-\partial M /\partial n$, $\partial\mu/\partial B$ can be expressed as the derivative of the magnetization $M$ per unit area with respect to the density.

We measured \dm{} in several high-mobility (100)-Si MOSFET structures similar to those used in Refs.\,\cite{simonyan,Prus2003,Pudalovsdh02} . Such a 2DES
possesses a twofold valley degeneracy in addition to its spin degeneracy. The in-plane magnetic field was modulated at frequency $\omega/2\pi=4-12$~Hz with amplitudes up to $40$\,mT. Measurements were performed over a wide range of
temperatures ($0.4-20$K), and in magnetic fields up to 9 T.

To get an insight into the expected behavior of \dm{} as a function of external parameters, let us briefly review
two limiting cases (i) of very high and (ii) of very low densities;  in both cases electrons can be viewed as noninteracting.

(i) For high densities, deep in the metallic regime, the ratio between interaction and the Fermi energies is small. One
expects to get the density-independent Pauli spin susceptibility at a magnetic field below the field of full spin
polarization $2n/(\nu g\mu_B)$, where $\nu$ is the density of states, and as a result $\partial M/\partial n=0$.

(ii) For very low densities, each electron is localized in its own potential well. Interactions between electrons are small; hence, the electron spins should be polarized independently: $\partial M/\partial n=\mu_B\tanh(b)$, where $b=\mu_B B/k_B T$ is a normalized magnetic field\,\footnote{The effective $g$-factor in the Si conduction band is 2}.

\begin{figure}
\vskip.05in
\begin{center}
\centerline{\psfig{figure=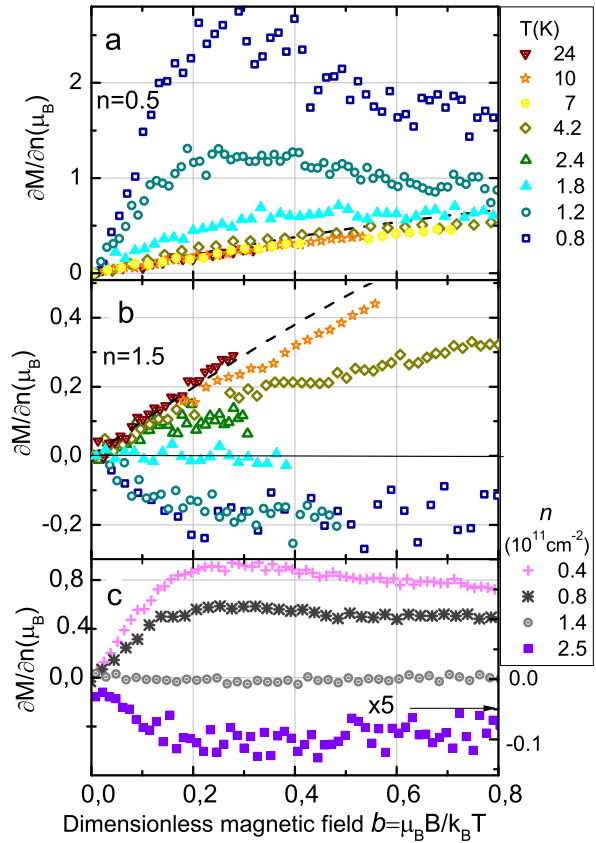, width=230pt}}
\caption{Panel (a):  $\partial M/\partial n$ vs normalized magnetic field  $b=\mu_B B/k_BT$ at $n=0.5\times 10^{11}{\rm cm^{-2}}$. We subtracted the diamagnetic contribution estimated from the high-temperature  data, $\sim 0.04 \mu_B$ per tesla \cite{Reznikov_2010}. Panel (b): the same as panel (a) at  $n=1.5\times 10^{11}{\rm cm^{-2}}$; the subtracted diamagnetic contribution is $\approx 0.035 \mu_B$ per tesla. Dashed lines in (a) and (b) show  $\partial M/\partial n$ for localized spins $1/2$. Panel (c) $\partial M/\partial n$ vs  $b$ at different densities at $T=1.8$K. Note that $\partial M/\partial n(b)$ becomes nonlinear at the density and temperature independent $b^*$.}
\label{dmdn_vs_B}
\end{center}\vskip -0.5in
\end{figure}

In Fig.\,\ref{dmdn_vs_B}(a) we present the magnetic field dependence of \dm{} measured deep in the insulating phase at $n=5\times 10^{10}{\rm cm^{-2}}$  ($n_c\approx 8.5\times 10^{10}{\rm cm^{-2}}$ for this sample). The results are shown after subtraction of the diamagnetic contribution\,\footnote{At temperatures exceeding correlation and Fermi energies we expect the spin susceptibility to be a function of the normalized magnetic field $b$. We therefore chose the diamagnetic contribution by making the data points at $T>7$\,K collapse onto a single curve. This results in diamagnetic contributions of $0.04 {\rm \mu_B}$ and $0.035 {\rm \mu_B}$ per tesla for densities $n=5\times 10^{10}{\rm cm^{-2}}$ and $n=1.5\times 10^{11}{\rm cm^{-2}}$ respectively, in agreement with the numerical estimations\,\cite{Reznikov_2010}.}; the subtraction does not affect the low-temperature results in any significant way. For temperatures above $\approx4.2$\,K, \dm{} is consistent with the expected dependence $\partial M/\partial n=\mu_B\tanh(b)$ for individual spins. However, as temperature decreases, (i) the low-field slope of \dm{} vs. $b$ becomes much steeper than the one expected for an independent spin $1/2$; (ii) \dm{} vs $b$ becomes nonlinear; and (iii) at low temperatures \dm{} reaches a maximum  at $b\approx b^*$\,\footnote{we define $b^*$ as the field at which the curvature \dm{} is maximal}; this maximum significantly exceeds $\mu_B$. The fact that $\partial M/\partial n>\mu_B$ means that an electron added to the system not only aligns its spin with the field but also promotes spin alignment of neighboring electrons. This is the ``smoking gun'' evidence for the ferromagnetic interaction between spins. Indeed, all these observations can be simulated even within the mean field approximation, if one assumes that ferromagnetic interaction constant grows with density, see the Supplemental  Material.

At higher densities, well inside the metallic phase, e.g. at $n=1.5\times 10^{11}{\rm cm^{-2}}$ shown in Fig.\,\ref{dmdn_vs_B}(b), low temperature \dm{} changes sign. Note that negative \dm{} is expected in the metallic phase, since increase in density reduces interaction and therefore polarization of the 2DES.  Thermal fluctuations also suppress magnetic ordering; therefore \dm{} becomes less negative with temperature and, at temperatures exceeding the Fermi energy (about 10K at $n=1.5\times 10^{11}{\rm cm^{-2}}$ ) approaches the dependence expected for noninteracting electrons.

Most importantly, the characteristic normalized magnetic field $b^*$ is almost constant over a wide range of densities [See Fig.\,\ref{dmdn_vs_B}(c)], from $\sim n_c/2$ to $\sim 3n_c$, thus excluding any possibility of a density-driven quantum phase transition into a homogenous ferromagnetic state. Rather it is reminiscent of the behavior of a large spin $J=(1/2) (1/b^*) \sim 2$ system.

Our results cannot be attributed to localized spins solely, whose interaction is known to be antiferromagnetic\,\cite{BhatReview}. A minimal model that would explain them is a two-phase state, consisting of electron droplets with a typical spin of 2 and itinerant electrons. As the density increases in the
insulating phase, electrons join already existing droplets or populate new ones, which leads to a positive \dm{}. For even higher densities these droplets start to coexist with  itinerant electrons; the addition of an electron to the system increases screening and therefore depopulates the droplets, which leads to negative \dm{} of the same order of magnitude and the same characteristic magnetic field scale $b^*$ as in the insulating phase.

\begin{figure}
\vskip.05in
\begin{center}
\centerline{\psfig{figure=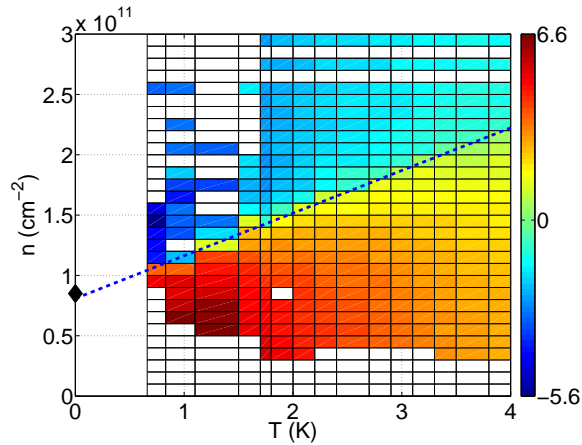, width=230pt}}
\caption{Phase diagram of \dchi{} represented in colors for each temperature and density, in units of ($\mu_B/T$). Dashed blue line represents the density, $n_{0}$, at which  \dchi{} is zero. Black $\blacklozenge$ - depicts the MIT critical density, $n_c$, known the from transport measurement.}
\label{phasediagram}
\end{center}
\end{figure}
\begin{figure}
\vskip.05in
\begin{center}
\centerline{\psfig{figure=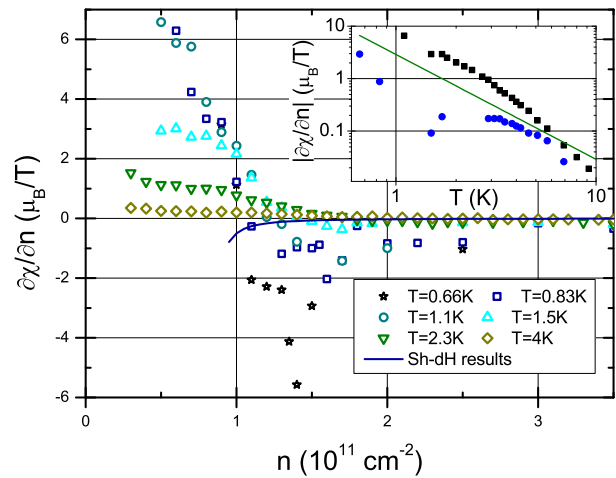, width=230pt}}
\caption{$\partial\chi/\partial n$ for different temperatures. Solid line represents $\partial\chi/\partial n$ extracted from the Sh-dH results\,\cite{Pudalovsdh02}. Inset shows temperature dependence of $|\partial\chi/\partial n|$ for two densities: ($\blacksquare$)$n=5\times 10^{10}{\rm cm^{-2}}$, and ($\bullet $)$n=1.5\times 10^{11}{\rm cm^{-2}}$ with data points in the vicinity of $\partial\chi/\partial n=0$ omitted; the solid line is $|\partial\chi/\partial n|\propto T^{-2}$}
\label{dchidn_vs_n}
\end{center}\vskip -0.5in
\end{figure}

The density at which \dm{} changes sign lies in the metallic phase and is temperature dependent; as $T\rightarrow 0$, it approaches the MIT critical density, $n_c$. In order to show this, let us focus on the low-field slope of \dm{} in Fig.\,\ref{dmdn_vs_B}, $\partial \chi/\partial n=\partial^2M/\partial n \partial B$. In Fig.\,\ref{phasediagram}, we plot a color map of \dchi{} for different temperatures and densities. The density $n_0$ at which $\chi$ is maximal ($\partial \chi/\partial n=0$) extrapolates linearly to $n_c$  with decreasing temperature. The coincidence of  the magnetic and transport critical density values,  $n_0$ and $n_c$, at $T=0$ suggests that the two phenomena: the formation of easily magnetized droplets and the MIT are intimately interrelated.

One would expect Curie ($1/T$) temperature dependence for the spin susceptibility of a droplet. In contrast, we found the temperature dependence of $|\partial \chi/\partial n |$ to be closer to $1/T^2$ both in the insulating phase and in the metallic phase away from the $\partial \chi/\partial n =0$ region; see inset in Fig.\,\ref{dchidn_vs_n}. This indicates that the number of droplets increases as $\propto 1/T$ in the temperature range of our measurements.

In Fig.\,\ref{dchidn_vs_n}, we present cross sections of the data of Fig.\,\ref{phasediagram} at several temperatures. Deep in the insulating phase (low densities),  \dchi{} is positive and for low enough temperatures far exceeds the Curie value for independent spins. As density increases, \dchi{} drops, changes sign, reaches a minimum, and eventually goes to zero, as expected for Pauli susceptibility of noninteracting electrons, which is density independent. As temperature decreases, the transition from low-density positive $\partial\chi/\partial n$ to the high-density negative one becomes steeper, indicating development of a sharp cusp in $\chi$ in the vicinity of $n_c$ at zero temperature. This behavior resembles the sharp drop of $dR/dT$ with density in the vicinity of the MIT.

Note that  transport measurements, both Shubnikov--de Haas\,\cite{Pudalovsdh02} (Sh-dH) and in-plane magnetoresistance\,\cite{SKDK,VZ2}, also found decrease in $\chi$ with density in the metallic phase. However, the magnitude of \dchi{} extracted from these measurements is much smaller than that measured with the recharging method, see solid line in Fig.\,\ref{dchidn_vs_n}. This discrepancy far exceeds any possible uncertainty in our experiment. We believe it is a result of the fundamental difference between the physical quantities provided by thermodynamic and transport methods: the thermodynamic method yields an average magnetization of all the electrons that are capable of recharging within the 80\,ms field modulation period. In contrast, the transport is influenced mostly by delocalized electrons having the picosecond-scale relaxation time.

In conclusion, we presented experimental evidence for the existence of spin droplets in high-mobility 2D electron layers in Si MOSFET samples on both sides of the MIT. The absence of similar behavior in low-mobility samples, for which the MIT is much less pronounced and occurs at much higher critical density $n_c \approx 3 \times 10^{11}$\,cm$^{-2}$, emphasizes the importance of interactions for the spin-droplet formation. A typical total spin $J=1/2b^*\approx2$ of a droplet, estimated above coincides with the value predicted within the theory \cite{Narozhny2000} for quite a realistic interaction constant value $F_0^a =-0.5$, typical of this regime\,\cite{Klimov08}. The minimal size of such a droplet can be estimated as $\sqrt{2J/n}$. This size, $\approx 100$\,nm for $n=5\times 10^{10}{\rm cm^{-2}}$, is comparable with the gate oxide thickness of 200\,nm, which screens the potential fluctuations and, hence, sets their spatial scale. Our results are in qualitative agreement with numerical calculations which predict enhancement of spin polarizability by disorder\,\cite{Fleury2010}. The observations of spin droplets in the metallic phase are in line with the expectation of stochastically formed multispin fluctuations in Fermi-liquid\,\cite{Narozhny2000}, in the insulating phase with the expectations for spontaneous magnetization of a disordered quantum dot\,\cite{AndreevKamenev,Shep2001}.

Our results suggest a picture of a nonuniform state in which easily polarizable electron droplets coexist with Fermi liquid over a wide density range. The spatial order (if any) of electrons within spin droplets remains unknown; these might be e.g. Wigner crystallites, or even more sophisticatedly ordered droplet phases\,\cite{Spivak2004}. However, one needs to explain why these two coexisting subsystems interact so weakly. Indeed, $b^*$ does not manifest itself in transport measurements, and there is no saturation of phase breaking time at low temperatures due to scattering by the spin droplets\,\cite{Klimov08}, as expected\,\cite{aleiner2001}. A plausible explanation for this would be the large size and collective nature of the droplets, which suppress electron-droplet scattering with spin flip.

The existence of these droplets can be checked in several ways; the most direct one would be magnetic force microscopy, somewhat analogous to measurements of local compressibility in 2DES\,\cite{Yacoby1999}. Polarization of photoluminescence signal serves as a local probe for the spin polarization.  The temperature dependence of ESR signal may provide information on the typical magnetic moment and the signal width on the spin relaxation rates. Knight shift measured on $^{29}$Si can be used for the same purpose.

We thank I.\,S.\,Burmistrov and D.\,Podolsky for fruitful discussion. The work at LPI was supported by RFBR, programs of the RAS and the Russian Ministry of Education and Sciences and utilized the facilities of the LPI Shared Facility Center. The work at Technion was supported by the Israel Science Foundation, the United States-Israel Binational Science Foundation, and the Russell Berrie Nanotechnology Institute.

\appendix
\section{Supplemental Material}

\begin{figure}
\begin{center}
\centerline{\psfig{figure=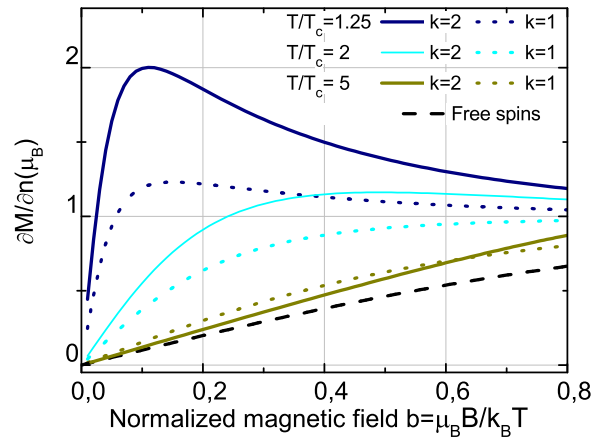, width=230pt}}
\caption{ \dm{} calculated within the mean field approximation. We assumed $T_c \propto n^k$; we show two sets of curves, solid lines for $k=2$, and dotted lines for $k=1$. Note similarity with the data presented in Fig.1a of the main text.}
\label{Fig:MF}
\end{center}\vskip -0.5in
\end{figure}

The simplest framework to include ferromagnetic interactions would be the mean field approximation, in which dimensionless magnetization per electron, $\tilde m=m/n\mu_B$, is given by
\be
\label{Eq:mean}
\tilde m = \tanh (b+\tilde m/\tilde t)
\ee
where $\tilde t=T/T_c$, and density-dependent parameter $T_c>0$ characterizes the interaction strength. We shall consider only $T>T_c$, since this model implies a ferromagnetic instability at $T=T_c$, forbidden in a 2D system with continuous symmetry.

We solved Eq.\,\ref{Eq:mean} numerically to get $\tilde m$ as a function of $b$ and $\tilde t$. The value of \dm{} can be expressed through $\tilde m$ and  depends on an additional parameter, $\partial T_c/\partial n$:

\be \label{Eq:dM}\frac{1}{\mu_B}\frac{\partial M}{\partial n}=\tilde m+n\frac{\partial \tilde m}{\partial n}=\tilde m-n\tilde t\frac{\partial \tilde m}{\partial \tilde t}
\frac{\partial T_c}{\partial n}
\ee

In the insulating phase (low densities) the interaction strength $T_c$ is expected to rise with $n$ from zero at $n=0$, due to increasing overlap between the localized electrons' wave functions. In the metallic phase (high densities) $T_c$ should decrease with density due to screening. Positive $\partial T_c/\partial n$ is precisely the condition required to reproduce the peak in ${\partial M}/{\partial n}(b)$. As an example we choose $T_c \propto n^k$; the results are qualitatively similar for $k=1$ and 2, see Fig.\,\ref{Fig:MF}. We stress that the mean field \dm{} qualitatively captures all the features seen in Fig.\,1a, including seemingly strange  ${\partial M}/{\partial n}>\mu_B$, and thus supports the claim of ferromagnetic interaction between spins in our system.

\end{document}